\documentclass[journal]{IEEEtran}

\usepackage{lineno,hyperref}
\usepackage{amsmath,amssymb}

\usepackage{booktabs}
\usepackage{makecell}
\usepackage{multirow}
\usepackage{graphicx}
\usepackage{multirow}
\usepackage{tabu,bm}
\usepackage{color}
\usepackage{subfigure}
\usepackage{lipsum,mwe,cuted}
\usepackage{booktabs}
\usepackage{stfloats}
\hypersetup{hidelinks}

\usepackage{algorithmic}
\usepackage[linesnumbered,ruled,vlined]{algorithm2e}
\usepackage{url}

\usepackage{cite}

\ifCLASSINFOpdf
\else
\fi
\begin{document}
%
\title{Hybrid Digital-Analog Semantic Communications}
%
%
%

\author{Huiqiang Xie,~\IEEEmembership{Member,~IEEE,}
        Zhijin Qin,~\IEEEmembership{Senior Member,~IEEE,} \\
        Zhu Han,~\IEEEmembership{Fellow,~IEEE,}
        and Khaled B. Letaief,~\IEEEmembership{Fellow,~IEEE}
\thanks{Huiqiang Xie is with College of Information Science and Technology, Jinan University, Guangzhou, China (e-mail: huiqiangxie@jnu.edu.cn).  }
\thanks{Zhijin Qin is with the Department of Electronic Engineering, Tsinghua University, Beijing, China, and also is with the State Key Laboratory of Space Network and Communications, Beijing, China. (e-mail: qinzhijin@tsinghua.edu.cn).  (\textit{Zhijin Qin is the corresponding author.}) }
\thanks{Zhu Han is with the Department of Electrical and Computer Engineering,
University of Houston, Houston, USA (e-mail: zhan2@uh.edu)). } 
\thanks{Khaled B. Letaief is with the Department of Electronic and Computer Engineering, The Hong Kong University of Science and Technology, Hong Kong, China (email: eekhaled@ust.hk).).  }
}


\maketitle

\begin{abstract}
Digital and analog semantic communications (SemCom) face inherent limitations such as data security concerns in analog SemCom, as well as leveling-off and cliff-edge effects in digital SemCom. In order to overcome these challenges, we propose a novel SemCom framework and a corresponding system called HDA-DeepSC, which leverages a hybrid digital-analog approach for multimedia transmission. This is achieved through the introduction of analog-digital allocation and fusion modules. To strike a balance between data rate and distortion, we design new loss functions that take into account long-distance dependencies in the semantic distortion constraint, essential information recovery in the channel distortion constraint, and optimal bit stream generation in the rate constraint. Additionally, we propose denoising diffusion-based signal detection techniques, which involve carefully designed variance schedules and sampling algorithms to refine transmitted signals. Through extensive numerical experiments, we will demonstrate that HDA-DeepSC exhibits robustness to channel variations and is capable of supporting various communication scenarios. Our proposed framework outperforms existing benchmarks in terms of peak signal-to-noise ratio and multi-scale structural similarity, showcasing its superiority in semantic communication quality.
\end{abstract}


\IEEEpeerreviewmaketitle

\section{Introduction}

As mobile communication systems transition from the fifth generation (5G) to the sixth generation (6G), there is a need to address the evolving requirements of seamlessly integrating virtual/augmented reality, remote control robots, and ubiquitous connected intelligence. To meet these demands, target key performance indicators \cite{WangYGZLZWHCHTLRTZSPH23} have been proposed, aiming to ensure the seamless integration of these advanced technologies in the next generation of mobile communication systems, e.g., $10^7$~devices/$\text{km}^2$ for connectivity, 60 b/s/Hz for spectral efficiency, and 100 us for end-to-end latency.  To materialize this vision, semantic communications \cite{qin01389} have been envisioned as one of the potential technologies due to the low semantic errors, high spectral efficiency, and high transmission rates. By exchanging semantic information at both ends, semantic communications can reconstruct sources or directly perform tasks with the tolerance of transmission errors. According to the communication paradigm, semantic communications can be categorized into two categories: \textit{analog semantic communications} and \textit{digital semantic communications}.

\textit{Analog semantic communications} \cite{Xie9398576, Yi10318078, Weng10038754, Grassucci10447612, Han9953316, Dai9791398, Zhang10431795, Wu10500305, Wang9953110} convey the semantic information using continuous signals, which takes advantage of deep learning (DL) to design end-to-end systems and maps the source to the non-fixed-size constellations directly. There exist many works for different modal data transmission. Xie \textit{et al.}~\cite{Xie9398576} have developed a DL based semantic communication system, named DeepSC, for text transmission, in which the sentences are mapped to the embedding vectors and then transformed to the learned non-fixed-size constellation points.  Yi \textit{et al.}~\cite{Yi10318078} introduced the explicit knowledge base to the DeepSC as the side information and integrated the knowledge base into the end-to-end optimization, such that achieves the higher bilingual evaluation understudy (BLEU) score at the low signal-to-noise ratio (SNR) regions. Weng \textit{et al.}~\cite{Weng10038754} have proposed an end-to-end semantic communication system for speech recognition and speech synthesis tasks, named DeepSC-ST. The speech signals are processed by the DeepSC-ST and output the continuous constellation points at the transmitter. Grassucci \textit{et al.}~\cite{Grassucci10447612} have designed a generative audio semantic communication framework, which transmits the continuous embedding vectors to generate the audios at the receiver. Dai \textit{et al.} \cite{Dai9791398} have investigated the end-to-end image transmission problem, in which the image is non-linearly transformed into continuous signals with different lengths. Wu \textit{et al.} \cite{Wu10500305} have investigated the end-to-end image transmission for multiple-inputs multiple-outputs (MIMO) channels. Similarly, the images are converted into continuous semantic features and adaptively assigned to different subchannels based on the channel state information (CSI). Wang \textit{et al.} \cite{Wang9953110} have proposed a video semantic communication system, in which the semantic features of frames are extracted into continuous signals and transmitted using analog communication methods.

The continuous signals in analog semantic communications have two \textit{benefits}. One is to allow gradient propagation and enable end-to-end optimization. The other is that the continuous signals have a high degree of freedom that provides the smoothness performance optimization varying from channel conditions, enabling better robustness in the low SNR regimes. However, continuous signals also have flaws. The commercial encryption algorithms are designed for discrete signals, e.g., bit streams, raising concerns about the data security of continuous signal-based systems.  Besides, in some scenarios that require accurate transmission at the bit level, analog semantic communications cannot meet the requirement due to the approximately infinite candidate sets in continuous signals. Therefore, digital semantic communications have attracted the attention of researchers.

\textit{Digital semantic communications} \cite{Tung9998051, Bo10495330, He10327757, Guo10200355, liu2024ofdm, Fu10065571, Hu10101778, Gao10251411, huang2024d} transmit semantic information in the type of discrete signal, which maps the source to bit streams or fixed-size constellations. Tung \textit{et al.}~\cite{Tung9998051} have proposed the quantized joint source-channel coding for image transmission, named DeepJSCC-Q, by mapping the continuous signals to the close points in the fixed-size constellations to be compatible with some protocols. Similarly, Bo \textit{et al.}~\cite{Bo10495330} improved the quantized joint source-channel coding by learning transition probability from source data to discrete constellation symbols, in which the Gumbel-Max sampling is employed to sample the constellation points from the learned transition probability so that avoiding the non-differentiable quantization.  Guo \textit{et al.}~\cite{Guo10200355} quantized the semantic information with the learnable non-linear scalar quantizer, which learns to adopt dynamic quantization levels for different semantic values. Fu \textit{et al.}~\cite{Fu10065571} have proposed the vector quantized semantic communication system, in which the semantic vectors are quantized into bit streams with the learnable vector quantizer and transmitted with the digital channel codings and modulations. Gao \textit{et al.}~\cite{Gao10251411} have developed an adaptive modulation and retransmission scheme by deriving the relationship between bit-error-rate and the task performance, in which the semantic information is quantized into fixed-length bit streams. Huang \textit{et al.}~\cite{huang2024d} have proposed an iterative training algorithm for digital semantic communications, in which the deep source codec are trained according to the chosen channel coding rate. 

The above works on digital semantic communication achieve accurate transmission at the bit or symbol level and part of the works can apply the encryption algorithms to encrypt the bit streams. However, digital semantic communication systems introduce unavoidable quantization errors due to the process of quantizing continuous signals to discrete signals, which introduces the \textit{leveling-off effect}. That is, the quality of the decoded source signal is limited because of the quantization errors. Besides, digital semantic communications experience the \textit{cliff-edge effect} varying from different channel conditions, which usually results in a drastic degradation in performance at lower SNRs. Therefore, it is imperative to adopt a new semantic communication paradigm that can address the limitations of both analog and digital semantic communications. This paradigm should enhance data security and mitigate the leveling-off and cliff-edge effects. However, designing such a semantic communication system poses several challenges that need to be overcome, namely,
\begin{enumerate}
    \item[\textit{Q1}:] \textit{How to enhance data security and alleviate the leveling-off and cliff-edge effects?}
    \item[\textit{Q2}:] \textit{How it be compatible with purely analog and digital semantic communication?}
    \item[\textit{Q3}:] \textit{How to support the various communication environments, e.g., the wide bandwidth scenario or the weak communication scenario?}
\end{enumerate}

The concept of hybrid digital-analog (HDA) joint source-channel codes \cite{Mittal995544} was proposed by Mittal \textit{et al.} in 2002, which proves that HDA codes are capable of theoretically achieving the Shannon limit (theoretically optimum distortion) and a less severe leveling-off and cliff-edge effects. Since then, the HDA codes have attracted much attention from academics and industries \cite{Skoglund1661853, Rüngeler6882172, Köken7138625, Fujihashi9424996}. Skoglund \textit{et al.} \cite{Skoglund1661853} have proposed HDA codes for the bandwidth compression scenarios, and Köken \textit{et al.} \cite{Köken7138625} have analyzed the robustness of HDA codes with bandwidth mismatch. HDA transmission is also adopted in the Japanese and Canadian television signal transmission \cite{Hart2004}, where video and speech signals are transmitted by analog and digital transceivers, respectively. However, these works rely on linear transforms and ignore the semantic information behind data, which is unsuitable for non-linear semantic transmission. 

Inspired by the concept of HDA codes, we propose a novel framework called DL-based HDA semantic communication. This framework integrates the strengths of both analog and digital semantic communications to effectively tackle the challenges mentioned earlier. Firstly, the HDA semantic communication systems can improve data security and alleviate the leveling-off and cliff-edge effects by transmitting part information with the continuous signals in analog communications (\textit{Q1}). Besides, analog and digital semantic communications are special cases of HDA semantic communications. By controlling the ratio between analog and digital components, the HDA semantic communications not only can be transformed into purely analog or digital semantic communications (\textit{Q2}) but also support the different communication scenarios (\textit{Q3}). The main contributions are summarized as follows:
\begin{itemize}
    \item A novel HDA semantic communication framework is proposed, which takes advantage of analog and digital semantic communications and addresses the limitations inherent in each. 
    \item Based on the HDA semantic communication framework, we propose an HDA semantic communication system, named HDA-DeepSC, for multimedia transmission, in which the new analog-digital allocation and fusion modules are proposed to generate the analog and digital components. Besides, the new loss functions are designed to capture the local and global information, alleviate the distortions from channels, and balance the source rate.
    \item To further improve the quality of the recovered images, we proposed a diffusion-based framework enhanced signal detection by designing the variance schedule and sampling algorithm.
    \item Based on extensive simulation results, the proposed HDA-DeepSC outperforms the conventional and DL-based communication systems and improves the system robustness at the low SNR regime.
\end{itemize}

The rest of this paper is organized as follows. The system model is introduced in Section \ref{sec-ii}.  The HDA semantic transmission is proposed in Section \ref{sec-iii}. Section \ref{sec-iv} details the proposed diffusion-based signal detection. Numerical results are presented in Section \ref{sec-v} to show the performance of the proposed frameworks. Finally, Section \ref{sec-vi} concludes this paper.

\textit{Notation}:  Bold-font variables denote matrices or vectors. $\mathbb{C}^{n \times m}$ and $\mathbb{R}^{n \times m}$  represent complex and real matrices of size $n\times m$, respectively. ${\cal CN}(\mu,\sigma^2)$ means circularly-symmetric complex Gaussian distribution with mean $\mu$ and covariance $\sigma^2$. ${\cal N}(\mu,\sigma^2)$ means Gaussian distribution with mean $\mu$ and covariance $\sigma^2$. ${\cal U}(a,b)$ means continuous  uniform distribution between $a$ and $b$. $(\cdot)^{*}$ denotes the conjugate operation.  ${\bm x}[k]$ represents the $k$-th element in the vector.

\section{System Model}\label{sec-ii}
\begin{figure*}[!t]
    \centering
    \includegraphics[width=150mm]{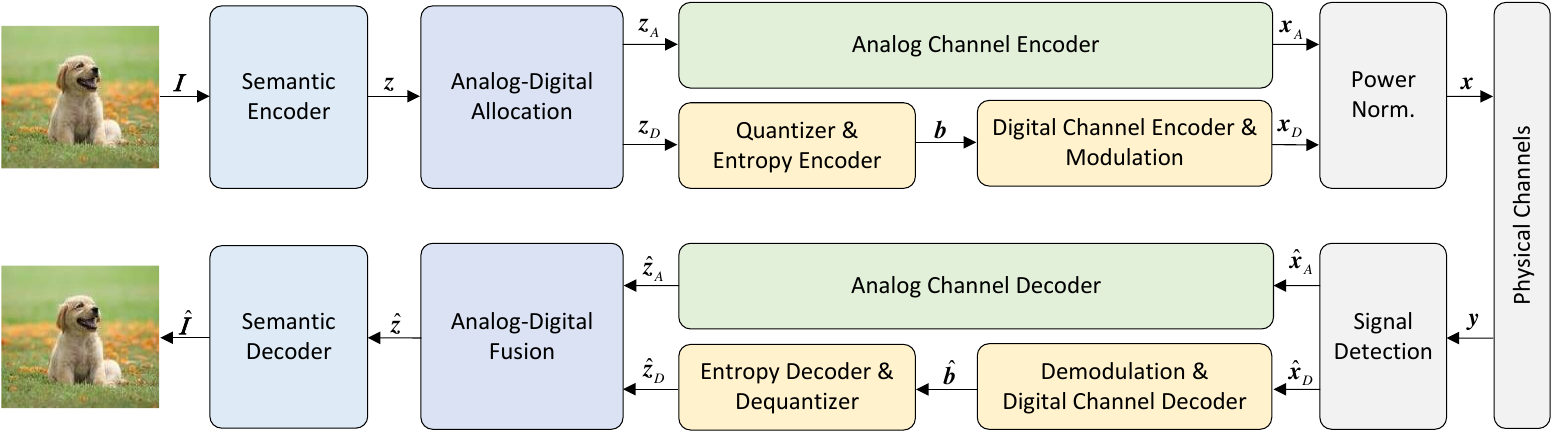}
    \caption{{The proposed hybrid digital-analog semantic communication framework.}}
    \label{fig:system-model}
\end{figure*}

As shown in Fig. \ref{fig:system-model}, we consider a single-input single-output (SISO) communication system, which aims to send multimedia over the air. The proposed HDA SemCom framework consists of the HDA transmitter, the wireless channel model, and the HDA receiver, which employs both digital semantic transmission and analog semantic transmission.

\subsection{The Hybrid Digital-Analog Transmitter}
The HDA transmitter consists of a semantic encoder that extracts the semantic information behind images, analog-digital allocation that allocates the semantic information for analog and digital transmission, and channel encoders that protect the information over the air.

Given an image, $\bm{I}\in {\mathbb R}^{3\times H \times W}$, where $H$ and $W$ are the height and width of the image. The semantic information can be extracted by
\begin{equation}
    \bm z = \mathcal{S}(\bm{I};{\bm \alpha}_t),
\end{equation}
where $\bm z \in {\mathbb R}^{M\times 1}$ is the semantic information and $\mathcal{S}(\cdot;{\bm \alpha}_t)$ is denoted as the semantic encoder with the parameter ${\bm \alpha}_t$. Then, $\bm z$ is split into two parts with analog-digital allocation module by
\begin{equation}\label{eq-2}
    [{\bm z}_{A}, {\bm z}_{D}] =\mathcal{A}(\bm z; {\bm \theta}_t),
\end{equation}
where ${\bm z}_{A}$ and  ${\bm z}_{D}$ are the semantic information transmitted by the analog transmitter and the digital transmitter, respectively. $\mathcal{A}(\cdot; {\bm \theta}_t)$ is analog-digital allocation with parameters ${\bm \theta}_t$.

\subsubsection{Analog Transmitter}
The encoded symbols for analog semantic transmission are represented as
\begin{equation}
    \bm{x}_{A} = \mathcal{C}_{A}(\bm{z}_{A};{\bm \beta}_t),
\end{equation}
where ${\bm x}_A \in \mathbb{C}^{L_A\times 1}$ is the encoded complex symbols and $\mathcal{C}_{A}(\cdot;{\bm \beta}_t)$ is denoted as the analog channel encoder with the parameter ${\bm \beta}_t$. 

\subsubsection{Digital Transmitter}

The entropy coding and quantizer will be employed firstly to convert ${\bm z}_{D}$ into bit streams by
\begin{equation}
    {\bm b} = \mathcal{E}(\mathcal{Q}({\bm z}_{D})),
\end{equation}
where ${\bm b}$ is the bit streams, $\mathcal{Q}(\cdot)$ and $\mathcal{E}(\cdot)$ are denoted as the quantizer and entropy encoder, respectively. Then, ${\bm b}$ is encoded with digital channel encoders (e.g., LDPC codes) and fixed-size constellations (e.g., 16-QAM) by
\begin{equation}
    {\bm x}_D = \mathcal{M}\left( \mathcal{C}_D({\bm b}) \right),
\end{equation}
where ${\bm x}_D \in \mathbb{C}^{L_D\times 1}$ is the encoded symbols, $\mathcal{M}(\cdot)$ represents the fixed-size modulation, and $\mathcal{C}_D(\cdot)$ is denoted as the digital channel encoder.

With the analog and digital symbols, the transmitted symbols are ${\bm x} = [{\bm x}_A, {\bm x}_D] \in   \mathbb{C}^{L\times 1}$, where $L=L_A + L_D$. The bandwidth compression ratio is defined as $\eta = \frac{L}{3\times H \times W}$.

\subsection{Wireless Channel Model}
When $\bm x$ is transmitted over the block fading channels, the received signal can be given by
\begin{equation}\label{eq-7}
    {\bm y} = {h} {\bm x} + {\bm n},
\end{equation}
where $ {h} $ is the channel coefficient that remains constant within a channel coherence time, $ {\bm n}$ is the additive white Gaussian noise (AWGN), in which ${\bm n} \sim \mathcal{CN}\left(0, \sigma_n^2{\bf I}_{{L}}\right)$.  For the Rayleigh fading channel, the channel coefficient follows ${h} \sim {\cal CN}\left(0, 1\right)$; for the Rician fading channel, it follows ${{ h}\sim \cal CN}\left(\mu_h,\sigma_h^2\right)$ with $\mu_h = \sqrt{r/(r+1)}$ and $\sigma_h = \sqrt{1/(r+1)}$,  where $r$ is the Rician coefficient. The SNR is defined as $\mathbb{E}({{{\left\| {\bm x}\right\|}^2}}) /\mathbb{E}({{{\left\| \bm n \right\|}^2}})$. 

\subsection{The Hybrid Digital-Analog Receiver}
The receiver comprises signal detection that estimates the transmitted symbols, a analog-digital fusion module that fuses the digital and analog semantic information, channel decoders that alleviate the distortions from the wireless channels, and a semantic decoder that recovers the images with the received semantic information.

With the least squares (LS) signal detection, the transmitted symbols can be estimated by 
\begin{equation}\label{eq-8}
    {\hat { \bm x}} = \frac{{{{ h}^*}}}{{{{| { h} |}^2}}}{\bm y} = {\bm x} + \frac{{{{ h}^*}}}{{{{| { h} |}^2}}}{\bm n},
\end{equation}
where ${\hat { \bm x}}=[{\hat { \bm x}}_A, {\hat { \bm x}}_D] $ represents the estimated symbols. We assume that ${ h}$ is the perfect CSI. After the signal detection, the semantic features are recovered by the analog and digital receivers, respectively.

\subsubsection{Analog Receiver}

The semantic features transmitted by analog communications are estimated by
\begin{equation}
    {\hat {\bm z}}_A = {\cal C}_A^{-1}({\hat { \bm x}}_A; {\bm \beta}_r),
\end{equation}
where ${\hat {\bm z}}_A $ is the estimated semantic features and  ${\cal C}_A^{-1}(\cdot; {\bm \beta}_r)$ is denoted as the analog channel decoder with parameter ${\bm \beta}_r$. 

\subsubsection{Digital Receiver}
For digital semantic transmission, the transmitted bit streams are recovered firstly by 
\begin{equation}
    {\hat {\bm b}} = \mathcal{C}_D^{-1}\left( \mathcal{M}^{-1}\left({\hat {\bm x}}_D\right) \right),
\end{equation}
where $\mathcal{C}_D^{-1}(\cdot)$ represents the digital channel decoder and $\mathcal{M}^{-1}(\cdot)$ is denoted as the fixed-size demodulation. Then, the semantic features transmitted with digital semantic transmission are recovered by
\begin{equation}
     {\hat {\bm z}}_D = \mathcal{Q}^{-1}(\mathcal{E}^{-1}( {\hat {\bm b}})),
\end{equation}
where $\mathcal{E}^{-1}(\cdot)$ and $\mathcal{Q}^{-1}(\cdot)$ are denoted as the entropy decoder and dequantizer, respectively.

With ${\hat {\bm z}}_A$ and ${\hat {\bm z}}_D$, the semantic features are fused by 
\begin{equation}
    {\hat {\bm z}} = {\mathcal{A}^{-1}\left({\hat {\bm z}}_A, {\hat {\bm z}}_D; {\bm \theta}_r\right)},
\end{equation}
where ${\hat {\bm z}}$ is the recovered semantic information and ${\mathcal{A}}^{-1}(\cdot;{\bm \theta}_r)$ is represented as the analog-digital fusion module with parameters ${\bm \theta}_r$. 

Finally, the transmitted image can be reconstructed by 
\begin{equation}
    {\hat {\bm I}} = \mathcal{S}^{-1}( {\hat {\bm z}};{\bm \alpha}_r),
\end{equation}
where $\mathcal{S}^{-1}( \cdot;{\bm \alpha}_r)$ represents the semantic decoder with parameter ${\bm \alpha}_r$.

\begin{figure*}[!t]
    \centering
    \includegraphics[width=150mm]{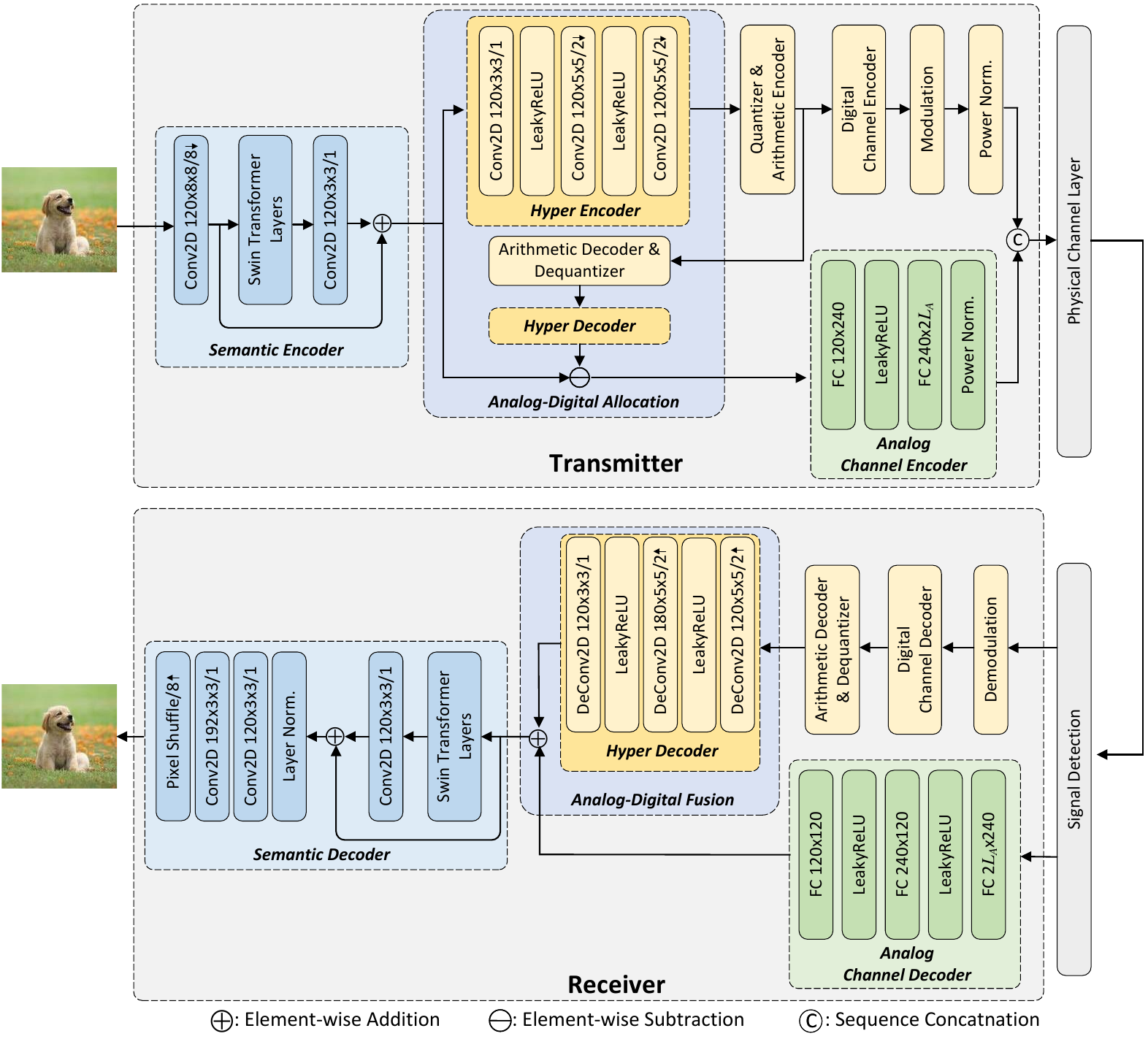}
    \caption{{The structure of the proposed hybrid digital-analog semantic communication system.}}
    \label{fig:had-deepsc}
\end{figure*} 

\section{Hybrid Digital-Analog Semantic Transmission}\label{sec-iii}

In this section, we design an HDA semantic communication system, named HDA-DeepSC, for heterogeneous wireless communication environments. Then, we develop the new loss function to train the HDA-DeepSC with the proposed training algorithm.

\subsection{Model Design}

The proposed HDA-DeepSC is shown in Fig. \ref{fig:had-deepsc}. The design of each module is detailed below.

\subsubsection{Semantic Codec}
The semantic encoder comprises a convolutional layer and a residual Swin Transformer block. The first convolutional layer projects the images into vector-shaped tokens, which are used as inputs to the residual Swin Transformer block in a permutation-invariant manner. Then, the residual Swin Transformer block consists of several Swin Transformer layers and a convolution layer, in which the Swin Transformer layer \cite{LiuL00W0LG21} originates from the Transformer and introduces the local attention and shifted window mechanism to improve the visual semantic understanding. Besides, a convolutional layer with spatially invariant filters in the residual block can enhance the translational equivariance. The residual connection allows for aggregation of the shallow and deep semantic features.

Similarly, the semantic decoder consists of the residual Swin Transformer block, convolutional layers, and pixel shuffle. The residual block is to enhance the visual semantic understanding. The residual connection provides a short connection from the semantic encoder to the semantic decoder, allowing the processing of reconstruction to fuse varying levels of features. The convolutional layers and pixel shuffle form the reconstruction module, in which the sub-pixel convolutional layer upsamples the feature and pixel shuffle reallocates the features to reconstruct the transmitted images.

\subsubsection{Analog-Digital Allocation and Fusion}

At the transmitter, the analog-digital allocation module transforms the original semantic information into essential and auxiliary semantic information. The essential part includes the basic information about the image, e.g., the high-frequency information, and needs to be delivered accurately and cryptographically. Therefore, the essential part is transmitted accurately by digital communication systems, in which the data encryption methods (e.g., symmetric cryptography and asymmetric cryptography) can be applied to encrypt the bit streams to guarantee the data security of the essential part. 

A hyper codec is proposed to extract the essential part of the original semantic information, which is given by
\begin{equation}
    {\bm z}_D = {\mathcal H}({\bm z}; {\bm \theta}_t),
\end{equation}
where ${\mathcal H}({\bm z}; {\bm \theta}_t)$ is denoted as the hyper encoder.  As shown in Fig. \ref{fig:had-deepsc}, the hyper encoder employs two convolutional layers to downsample the original semantic information, such that enables a larger receptive field and extracts the essential semantic information.

The auxiliary part helps improve the quality of the recovered image, which is transmitted by analog communication systems with the following benefits. Analog communication systems do not have a cliff effect and are suitable for optimizing systems in an end-to-end manner. To extract the auxiliary part, we first analyze the entropy of $\bm z$ conditioned on ${\tilde {\bm z}}$, $H\left( {{\bm z}\left| {\tilde {\bm z}} \right.} \right)$, which qualifies the uncertainty about  $\bm z$ when  ${\tilde {\bm z}}$ is known. In other words, it measures the remaining information of $\bm z$ when ${\tilde {\bm z}}$ is known. The lower bound of  $H\left( {{\bm z}\left| {\tilde {\bm z}} \right.} \right)$ is derived by
\begin{equation}\label{eq-15}
    \begin{aligned}
    H\left( {{\bm z}\left| {\tilde {\bm z}} \right.} \right) &= H\left( {{\bm z}, {\tilde {\bm z}}} \right) - H\left( {\tilde {\bm z}} \right)\\
     &\ge H\left( {\bm z} \right) - H\left( {\tilde {\bm z}} \right), \\
    \end{aligned}
\end{equation}
where the equals hold when $\tilde {\bm z}$ is close to $\bm z$. ${\tilde {\bm z}} = {\mathcal H}^{-1}\left({\mathcal{Q}}^{-1}\left({\mathcal{Q}}({\bm z}_D)\right); {\bm \theta}_r \right)$ is the recovered semantic information based on essential part without consideration of transmission errors. ${\mathcal H}^{-1}(\cdot;{\bm \theta}_r)$ is denoted as hyper decoder, where two convolutional layers are employed to upsample and recover the basic semantic information. 

By observing \eqref{eq-15}, we can obtain the remaining information of $\bm z$ when ${\tilde {\bm z}}$ is known, i.e., the auxiliary part, by
\begin{equation}
    {\bm z}_A = {\bm z} - {\tilde {\bm z}},
\end{equation}
where ${\bm z}_A$ is transmitted by analog communications.

At the receiver, the analog-digital fusion module is employed to obtain the fine semantic information by fusing the essential and auxiliary parts, which is given by
\begin{equation}
    {\hat {\bm z}} = {\mathcal{A}^{-1}\left({\hat {\bm z}}_A, {\hat {\bm z}}_D; {\bm \theta}_r\right)} = {\mathcal H}^{-1}({\hat {\bm z}}_D; {\bm \theta}_r) + {\hat {\bm z}}_A,
\end{equation}
where ${\mathcal H}^{-1}(\cdot;{\bm \theta}_r)$ shares the same weights with the hyper decoder in the transmitter. 

The design of analog-digital allocation and fusion can also be viewed as a coarse-to-fine processing. The digital and analog component transmits coarse and auxiliary semantic information about the basics and supplements of the image, respectively. The receiver fuses the coarse and auxiliary semantic information to obtain fine semantic information, which is used to recover the high-fidelity images.

\subsubsection{Digital Transceiver}
The quantizer module rounds elements of ${\bm z}_D$ to the nearest integer, ${\tilde {\bm z}}_D$. Then, the arithmetic coding converts ${\tilde {\bm z}}_D$ into bit streams, in which the arithmetic coding is one kind of entropy coding. The entropy coding requires the distribution of ${\tilde {\bm z}}_D$ in advance. Similarly to \cite{BalleMSHJ18}, we model  ${\tilde {\bm z}}_D$ using a non-parametric, fully factorized density model by
\begin{equation}\label{eq-18}
   p\left( {{\tilde {\bm z}}_D \left| {\bm \psi } \right.} \right) = \mathop \prod \limits_i {{p_{{{\tilde {\bm z}}_{D}[i]}\left| {{{\bm \psi }[i] }} \right.}}}\left( {{\bm \psi }[i] * {\mathcal U}\left( { - \frac{1}{2},\frac{1}{2}} \right)} \right)\left( {{\tilde {\bm z}}_D[i]} \right),
\end{equation}
where ${\bm \psi }[i]$ is the parameters of each univariate distribution ${p_{{\tilde {\bm z}}_D[i]\left| {{{\bm \psi }[i]}} \right.}}$. Like most cases, we model the quantization errors with the uniform distribution. Therefore, we convolve each non-parametric density with a standard uniform density to better match the prior of ${\tilde {\bm z}}_D$.

For digital channel codec and modulation, we adopt the adaptive modulation and coding for different SNRs.

\subsubsection{Analog Transeiver}
\textcolor{black}{The analog channel codec aims to compress the semantic features and transmit them effectively over the air.} Similarly to the previous works \cite{Xie9830752}, the analog channel codec mainly employs the fully connected layers to transmit the semantic information due to global semantic information preservation.  \textcolor{black}{Compared with the convolutional neural network (CNN) layer to capture the local information, the dense layer is good at capturing global information and preserving the entire attributes, which follows the target of the analog channel codec. This can enhance the system's robustness to channel noise.}

\subsection{Loss Function Design}
The wireless multimedia transmission problem can be viewed as the classical rate-distortion optimization problem, which includes distortion and rate constraints. 

\subsubsection{Loss Function Design for Distortion Constraints}
The distortion constraint can be categorized into semantic and channel distortion constraints. For semantic distortion constraint, except for the pixel difference considered in most works, we further introduce the frequency difference of the images. The designed loss function for semantic distortion constraint is given by 
\begin{equation}\label{eq-19}
    {\mathcal L}_{\text{SD}} = {\mathbb E}\left[  {\| {{{\bm I}} - {\hat {\bm I}}} \|^2}  + \lambda_{\mathcal F}  | {{\mathcal F}\left( { {\bm I}} \right) - {\mathcal F}( {\hat {\bm I}} )} |\right],
\end{equation}
where $\lambda_{\mathcal F}$ is the weight and ${\mathcal F}(\cdot)$ represents the Fourier transform. The first item in \eqref{eq-19} refers to the pixel difference of the image, we assume that the pixels of the image follow the Gaussian distribution without loss of generality and thus employ the mean-square error (MSE) loss. The second item in \eqref{eq-19} refers to the frequency difference of the image, we consider the learning of long-range dependencies of the image and design the Fourier-based loss function. In detail, we map the images into the frequency domain and compare the difference between the original and transmitted images. The reasons behind the design can be summarized as

\begin{itemize}
    \item The MSE loss guides the neural networks to recover the local pixels of the images by comparing the pixel difference, which ignores the long-range dependencies of the image.
    \item The Fourier-based loss can help the neural network learn the long-range dependencies of the image. Because the same frequency in the frequency domain refers to the different pixels at the different positions of the image.
\end{itemize}

For the channel distortion constraint, we consider the distortions from channels and the transmission of essential information. The designed loss function is given by 
\begin{equation}\label{eq-20}
   {\mathcal L}_{\text{CD}} = {\mathbb E} \left[ \| {\bm z} - {\hat {\bm z}}\| \right] - I({\bm z}, {\hat {\bm z}}_D),
\end{equation}
where the first item minimizes the distortions from channels and the second item maximizes the mutual information between ${\bm z}$ and ${\hat {\bm z}}_D$ to make ${\hat {\bm z}}_D$ contains more information of ${\bm z}$. However, directly optimizing the $I({\bm z}, {\hat {\bm z}}_D)$ is hard. We derive the lower bound of $I({\bm z}, {\hat {\bm z}}_D)$ by
\begin{equation}
\begin{aligned}
I ( {{\bm z},{{\hat {\bm z}}_D}} )  &= \int {p\left( {{\bm z},{{\hat {\bm z}}_D}} \right)} \log \frac{{p\left( {{\bm z}\left| {{{\hat {\bm  z}}_D}} \right.} \right)}}{{p\left( {\bm z} \right)}}d{\bm z}d{{\hat {\bm z}}_D}\\
 &\ge \int {p\left( { {\bm z},{{\hat {\bm z}}_D}} \right)} \log q\left( {{\bm z}\left| {{{\hat {\bm z}}_D}, {\bm \theta}_r } \right.} \right)d{\bm z}d{{\hat {\bm z}}_D}  \\
 & \quad - \int {p\left( {\bm z} \right)} \log p\left( {\bm z} \right)dz\\
 &= \! \! \int \! \! {p\left( {\bm z} \right)p\left( {{{\hat {\bm z}}_D}\left| {\bm z} \right.} \right)} \log q\left( {{\bm z}\left| {{{\hat {\bm z}}_D}, {\bm \theta}_r} \right.} \right)\!d{\bm z}d{{\hat {\bm z}}_D} \! +\! H\left( {\bm z} \right)  \\
 &= {{\mathbb E}_{{\bm z} \sim p\left( {\bm z} \right)}}{{\mathbb E}_{{{\hat {\bm z}}_D} \sim p\left( {{{\hat {\bm z}}_D}\left| {\bm z} \right.} \right)}}\left[ {\log q\left( {{\bm z}\left| {{{\hat {\bm z}}_D}, {\bm \theta}_r} \right.} \right)} \right] + H\left( {\bm z} \right).
\end{aligned}
\end{equation}
where the inequation follows $\mathrm{KL}\left[p({\bm z}|{\hat {\bm z}_D}), q({\bm z}|{\hat {\bm z}_D}, {\bm \theta}_r)\right] \ge 0$,  in which $\mathrm{KL}[\cdot,\cdot]$ is the Kullback-Leibler (KL) divergence and $q\left( {{\bm z}\left| {{\hat {\bm z}_D},{\bm \theta}_r} \right.} \right)$ is the variational approximation of $p\left( {{\bm z}\left| {\hat {\bm z}_D} \right.} \right)$.

For the sake of argument, assume for a moment that the likelihood is given by
\begin{equation}
    q\left( {{\bm z}\left| {{\hat {\bm z}_D},{\bm \theta}_r} \right.} \right) = {\mathcal N}\left( {  { {\mathord{\buildrel{\lower3pt\hbox{$\scriptscriptstyle\frown$}} 
\over {\bm z}} },(2\lambda_{z})^{-1} {\bf{I}}}} \right),
\end{equation}
where ${\mathord{\buildrel{\lower3pt\hbox{$\scriptscriptstyle\frown$}} 
\over {\bm z}} }={\mathcal H}^{-1}({\hat {\bm z}}_D; {\bm \theta}_r)$. The log-likelihood then works out to be the squared difference between ${\bm z}$ and ${\mathord{\buildrel{\lower3pt\hbox{$\scriptscriptstyle\frown$}} 
\over {\bm z}} }$ weighted by $\lambda_{z}$. Then, the $I\left( {{\bm z},{{\hat {\bm z}}_D}} \right)$ can be rewritten as
\begin{equation}\label{eq-23}
    I\left( {{\bm z},{\hat {\bm z}_D}} \right) \ge -\lambda_{z} {\mathbb E} \left[ {\| {{\bm z} - {\mathord{\buildrel{\lower3pt\hbox{$\scriptscriptstyle\frown$}} 
\over {\bm z}} } }\|^2} \right] + H(z) + \text{constant}.
\end{equation}

Submitting \eqref{eq-23} into \eqref{eq-20} and omitting the constant, the ${\mathcal L}_{\text{CD}}$ can be written as
\begin{equation}\label{eq-24}
    {\mathcal L}_{\text{CD}} \approx {\mathbb E} \left[ \| {\bm z} - {\hat {\bm z}}\| \right] + \lambda_{z} {\mathbb E} \left[ {\| {{\bm z} - {\mathord{\buildrel{\lower3pt\hbox{$\scriptscriptstyle\frown$}} 
\over {\bm z}} } }\|^2} \right] - H(z).
\end{equation}
If we freeze the semantic codec during training, $H(z)$ can be technically dropped out from ${\mathcal L}_{\text{CD}}$. 
\begin{algorithm}[!t]
\caption{HDA-DeepSC Training Algorithm.}
\label{alg-1}
\SetKwInput{KwInput}{Input}                
\SetKwInput{KwInitia}{Initialization}
\SetKwInput{KwOutput}{Output}              
\SetKwInput{KwRet}{Return}
\DontPrintSemicolon

\SetKwFunction{FMain}{Main}
\SetKwFunction{FSE}{Train Semantic Codec}
\SetKwFunction{FCC}{Train Hybrid Transceiver}
\SetKwFunction{FWN}{Train Whole Network}

  \SetKwProg{Fn}{Function}{:}{}
  \Fn{\FSE{}}{
        \KwInput{ Sample ${\bm I}$ from dataset.}
        
   	 ${\bm z} = {{\mathcal S}\left( {{\bm I};{\bm \alpha}_t } \right)}$,\;
         ${\hat {\bm I}} = {{\mathcal S}^{-1}\left( {{\bm z};{\bm \alpha}_r } \right)}$,\;
   	 	Compute ${\mathcal L}_{\text{SD}}$ with \eqref{eq-19}, \;
         Train ${\bm \alpha }_t, {\bm \alpha }_r$ $\to$ Gradient descent with ${\mathcal L}_{\text{SD}}$.\;
        \KwRet{{${{\mathcal S}\left( \cdot; {\bm \alpha }_t\right)}$ and ${{\mathcal S}^{-1}\left( {\cdot; {\bm  \alpha }_r } \right)}$.}} 
  }

  \SetKwProg{Fn}{Function}{:}{}
  \Fn{\FCC{}}{
        \KwInput{Freeze semantic codec and sample Semantic features ${\bm z}$.}
        \textbf{Transmitter}:\;
            \quad // \texttt{Analog-Digital Allocation} \;
            \quad ${\bm z}_D = {\mathcal H}({\bm z}; {\bm \theta}_t)$, \;
            \quad ${\tilde {\bm z}}_D = {\bm z}_D + {\bm u}$, ${\bm u} \sim {\mathcal U\left(-\frac{1}{2}, \frac{1}{2}\right)}$,\;
            \quad ${\tilde {\bm z}} = {\mathcal H}^{-1}({\tilde {\bm z}}_D ; {\bm \theta}_r)$, \;
            \quad ${\bm z}_A = {\bm z} - {\tilde {\bm z}}$. \;
            \quad // \texttt{Digital Transmitter} \;
            \quad Transmit ${\tilde {\bm z}}_D$ error-free to avoid gradient disappear.\;
            \quad // \texttt{Analog Transmitter} \;
            \quad $\bm{x}_{A} = \mathcal{C}_{A}(\bm{z}_{A};{\bm \beta}_t)$, \;
   		\quad {Power normalization,} \;
   		\quad Transmit ${\bm x}_{A} $ over the air. \;
            
   	\textbf{Receiver}:\;
            \quad Receive ${\bm y}_A$ with \eqref{eq-7} and ${\tilde {\bm z}}_D$. \;
            \quad // \texttt{Digital Receiver} \;
            \quad ${\hat {\bm z}}_D = {\tilde {\bm z}}_D$, \;
            \quad $ {\mathord{\buildrel{\lower3pt\hbox{$\scriptscriptstyle\frown$}} 
\over {\bm z}}} = {\mathcal H}^{-1}({\hat {\bm z}}_D ; {\bm \theta}_r)$. \;
            \quad // \texttt{Analog Receiver} \;
   		\quad Signal detection by \eqref{eq-8} to get ${\hat {\bm x}}_A$,\;
   		\quad $ {\hat {\bm z}}_A = {{\mathcal C}_A^{-1}\left( {\hat {\bm x}_A;{\bm \beta}_r} \right)}$. \; 
            \quad // \texttt{Analog-Digital Fusion} \;
            \quad  $ {\hat {\bm z}} = {\hat {\bm z}}_A + {\mathord{\buildrel{\lower3pt\hbox{$\scriptscriptstyle\frown$}} 
\over {\bm z}}}$. \; 
   		Compute ${\mathcal L}_{\text{CD}} + \lambda_r{\mathcal L}_{\text{Rate}}$ with \eqref{eq-24} and \eqref{eq-25}.\;
        Train ${\bm \beta }_t, {\bm \beta }_r, {\bm \theta }_t, {\bm \theta }_r$ $\to$ Gradient descent with ${\mathcal L}_{\text{CD}} + \lambda_r{\mathcal L}_{\text{Rate}}$.\;
        \KwRet{${{\mathcal C}_A\left( {\cdot;{\bm \beta }_t} \right)}$, ${{\mathcal C}_A^{-1}\left( {\cdot;{\bm \beta }_r} \right)}$, ${{\mathcal H}\left( {\cdot;{\bm \theta }_t} \right)}$, and ${{\mathcal H}^{-1}\left( {\cdot;{\bm \theta }_r} \right)}$.}
  }
 
  \SetKwProg{Fn}{Function}{:}{}
  \Fn{\FWN{}}{
         \KwInput{ Sample $\bm I$ from dataset.}
        Repeat lines 2, 8-28, and 3 to get $\hat {\bm I}$.\;
  		Compute ${\mathcal L}_{\text{SD}} + \lambda_r{\mathcal L}_{\text{Rate}}$ with \eqref{eq-19} and \eqref{eq-25}.  \;
        Train ${\bm \alpha }_t, {\bm \alpha }_r$, ${\bm \beta }_t, {\bm \beta }_r, {\bm \theta }_t, {\bm \theta }_r$ $\to$ Gradient descent with ${\mathcal L}_{\text{SD}} + \lambda_r{\mathcal L}_{\text{Rate}}$.\;
        \KwRet{{The HDA-DeepSC.} }
  }
\end{algorithm}

\begin{algorithm}[!t]
\caption{HDA-DeepSC Inference Algorithm.}
\label{alg-2}
\SetKwInput{KwInput}{Input}                
\SetKwInput{KwInitia}{Initialization}
\SetKwInput{KwOutput}{Output}              
\SetKwInput{KwRet}{Return}
\DontPrintSemicolon

\SetKwFunction{FMain}{Main}
\SetKwFunction{FCC}{HDA-DeepSC Inference}

  \SetKwProg{Fn}{Function}{:}{}
  \Fn{\FCC{}}{
        \KwInput{Sample $\bm I$ from dataset.}
        \textbf{Transmitter}:\;
            \quad ${\bm z} = {{\mathcal S}\left( {{\bm I};{\bm \alpha}_t } \right)}$.\;
            \quad // \texttt{Analog-Digital Allocation} \;
            \quad ${\bm z}_D = {\mathcal H}({\bm z}; {\bm \theta}_t)$, \;
            \quad ${\tilde {\bm z}_D} = {\mathcal Q}({\bm z}_D)$,\;
            \quad ${\tilde {\bm z}} = {\mathcal H}^{-1}({\mathcal Q}^{-1}({\tilde {\bm z}}_D) ; {\bm \theta}_r)$, \;
            \quad ${\bm z}_A = {\bm z} - {\tilde {\bm z}}$. \;
            \quad // \texttt{Digital Transmitter} \;
            
            \quad $ {\bm b} =  {\mathcal E}({\tilde {\bm z}_D})$, \;
            \quad ${\bm x}_D = \mathcal{M}\left( \mathcal{C}_D({\bm b}) \right)$. \;
            \quad // \texttt{Analog Transmitter} \;
            
            \quad $\bm{x}_{A} = {\mathcal C}_{A}(\bm{z}_{A};{\bm \beta}_t)$, \;
   		\quad {Power Normalization,} \;
   		\quad Transmit ${\bm x} = [{\bm x}_{A}, {\bm x}_{D}]$  over the air.\;
     
   	\textbf{Receiver}:\;
            \quad Receive ${\bm y}$ with \eqref{eq-7}. \;
   		\quad Signal detection by \eqref{eq-8} to get ${\hat {\bm x}}_A$ and ${\hat {\bm x}}_D$.\;
            \quad // \texttt{Digital Receiver} \;
            \quad ${\hat {\bm b}} = \mathcal{C}_D^{-1}\left( \mathcal{M}^{-1}\left({\hat {\bm x}}_D\right) \right)$, \;
            \quad ${\hat {\bm z}}_D = \mathcal{Q}^{-1}(\mathcal{E}^{-1}( {\hat {\bm b}}))$, \;
            \quad $ {\mathord{\buildrel{\lower3pt\hbox{$\scriptscriptstyle\frown$}} 
\over {\bm z}}} = {\mathcal H}^{-1}({\hat {\bm z}}_D ; {\bm \theta}_r)$. \;
            \quad // \texttt{Analog Receiver} \;
   		\quad $ {\hat {\bm z}}_A = {{\mathcal C}_A^{-1}\left( {\hat {\bm x}_A;{\bm \beta}_r} \right)}$. \; 
     \quad // \texttt{Analog-Digital Fusion} \;
            \quad  $ {\hat {\bm z}} = {\hat {\bm z}}_A + {\mathord{\buildrel{\lower3pt\hbox{$\scriptscriptstyle\frown$}} 
\over {\bm z}}}$, \; 
\quad ${\hat {\bm I}} = {{\mathcal S}^{-1}\left( {{\hat {\bm z}};{\bm \alpha}_r } \right)}$.\;
   		
        \KwRet{$\hat {\bm I}$.}
  }

\end{algorithm}

\subsubsection{Loss Function Design for Rate Constraints}
For rate constraint, the analog transmitter designs the fixed-length output. Therefore, we consider the rate constraint for the digital transmitter, which is given  
\begin{equation}\label{eq-25}
    {\mathcal L}_{\text{Rate}} = {\mathbb E} \left[-\log(p\left( {{\tilde {\bm z}}_D \left| {\bm \psi } \right.} \right))  )\right],
\end{equation}
where $p\left( {{\tilde {\bm z}}_D \left| {\bm \psi } \right.} \right)$ is given in \eqref{eq-18}. By minimizing the rate constraint, we can optimize the distribution of ${\tilde {\bm z}}_D$ and reduce the number of bits generated by the arithmetic coding.

\subsection{Training Details}

The proposed training algorithm is shown in Algorithm \ref{alg-1}. We adopt three-stage training methods. The first stage is to train the semantic codec with the ${\mathcal L}_{\text{SD}}$, which enables effective semantic extraction. After the semantic codec finishes training, the second stage is to train the hybrid transceiver with ${\mathcal L}_{\text{CD}} + \lambda_r{\mathcal L}_{\text{Rate}}$, which aims to reduce the distortions from physical channels as well as the number of bit streams. We can drop out the $H(z)$ in ${\mathcal L}_{\text{CD}}$ since we freeze the semantic codec during training. The non-differentiable operations, e.g., the quantization, entropy coding, and modulation, will block the gradient back-propagation from receiver to transmitter. Therefore, we substitute additive uniform noise for the non-differentiable operations itself during training, i.e., ${\tilde {\bm z}}_D = {\bm z}_D + {\bm u}$ in line 10 of Algorithm \ref{alg-1}. Besides, we choose the error-free transmission for the ${\tilde {\bm z}}_D$ due to two factors, one is that the number of generated bit streams is much smaller than the conventional source coding, e.g., JPEG; another one is the accurate bit transmission characteristic of digital communication. Finally, we train the whole network with ${\mathcal L}_{\text{SD}} + \lambda_r{\mathcal L}_{\text{Rate}}$ to improve the quality of the recovered image and reduce the number of bit streams in an end-to-end manner, which converges to the global optimization.

When the whole network has been trained, we can employ the model to transmit the image wirelessly. The inference algorithm is presented in Algorithm \ref{alg-2}. We remove the additive uniform noise and replace it with the non-differentiable operations. 

\textcolor{black}{The three-stage training algorithm ensures that each stage can converge to the local optimum and avoids the mismatch of gradient descent. Besides, the approximate quantized noise helps avoid the disappearing gradient, which enables end-to-end training. Moreover, the inference algorithm indicates that the digital component can adopt the encryption algorithm to protect the digital bits and the adaptive modulation coding against channel distortions.}

\section{Diffusion Framework Enhanced Signal Detection}\label{sec-iv}

This section provides an overview of the de-noising diffusion framework and its background. Subsequently, we introduce a novel diffusion-based signal detection method called DiffSDNet. DiffSDNet is developed by incorporating a carefully designed variance schedule into the training and sampling algorithms.

\subsection{De-noising Diffusion Framework}
Given a random noise as input, the denoising diffusion framework \cite{HoJA20} models the generative processing through multiple de-noising steps. Each step iteratively enhances the generative results by removing the predicted noise, akin to Langevin dynamics. The de-noising diffusion framework is divided into forward process and reverse process.

\subsubsection{Forward Process}
The forward process is fixed to a Markov chain with $T$ steps that gradually adds Gaussian noise to the data according to a variance schedule ${\gamma_1}, \cdots, {\gamma_T}$, which is given by
\begin{equation}\label{eq-26}
    {{\bm x}^{(0)}} \to {{\bm x}^{(1)}} \to {{\bm x}^{(2)}} \to  \cdots  \to {{\bm x}^{(T - 1)}} \to {{\bm x}^{(T)}},
\end{equation}
where ${{\bm x}^{(0)}}$ is the input information, $p\left( {{{\bm x}^{(t)}}\left| {\bm x}^{(t - 1)} \right.} \right) = {\mathcal N} \left(  {\sqrt {1 - {\gamma^{(t)}}} {{\bm x}^{(t - 1)}},{\gamma^{(t)}}{\bf I}} \right)$, and $p({\bm x}^{(T)})$ is modeled with ${\mathcal N}({\bf 0}, {\bf I})$.  Due to the reparameterization of normal distribution, ${{\bm x}^{(t)}}$ can be represented as
\begin{equation}\label{eq-27}
    \begin{aligned}
        {{\bm x}^{(t)}} &= {\sqrt {1 - \gamma^{(t)}} } {\bm x}^{(t-1)} + {\sqrt {\gamma^{(t)}} } {\bm \epsilon}^{(t)} \\
        &= \sqrt{1 - {{\left( {{{\bar \gamma }^{(t)}}} \right)}^2} }  {\bm x}^{(0)} +  {{\bar \gamma }^{(t)}} {\bar {\bm \epsilon}}^{(t)},
    \end{aligned}
\end{equation}
where ${\bar {\bm \epsilon}}^{(t)} \sim {\mathcal N}({\bf 0},{\bf I})$ and $  {{{\bar \gamma }^{(t)}}}  = \sqrt{ {1 - \mathop \prod\nolimits_{t = 1} ( {1 - {\gamma ^{(t)}}} )  }} $. Observe \eqref{eq-27}, the forward process recurrently adds the Gaussian noise step by step to make ${{\bm x}^{(0)}}$ approach the normal distribution, which can be viewed as the encoding processing without learnable parameters.

\subsubsection{Reverse Process}
The reverse process is also defined as a Markov chain with $T$ steps starting at ${{\bm x}^{(T)}}$, which is given by
\begin{equation}\label{eq-28}
    {{\bm x}^{(T)}} \to {{\bm x}^{(T - 1)}} \to {{\bm x}^{(T - 2)}} \to  \cdots  \to {{\bm x}^{(1)}} \to {{\bm x}^{(0)}},
\end{equation}
where $q\left( { {\bm x}^{(t - 1)}\left| {{{\bm x}^{(t)}}} \right.} \right) = {\mathcal N}\left( { {\mu \left( { {\bm x}^{(t)}; {\bm \omega} } \right), \sigma^{(t)} {\bf I}} } \right)$. The reverse process generates the ${{\bm x}^{(t - 1)}}$ based on ${{\bm x}^{(t)}}$, in which the mean of ${{\bm x}^{(t - 1)}}$ is modeled with neural network with the ${{\bm x}^{(t)}}$ as input.

From \eqref{eq-27}, we can observe that ${\bm x}^{(t - 1)}$ can be predicted with ${\bm x}^{(t)}$ and ${\bm x}^{(0)}$ by removing the added noise. Therefore, ${\mu \left( { {\bm x}^{(t)}; {\bm \omega} } \right)}$ can be modeled as
\begin{equation}\label{eq-29}
    \mu ( { {\bm x}^{(t)}; {\bm \omega} } ) = \frac{1}{{\sqrt {1 - {\gamma^{(t)}}} }}\left( {{\bm x}^{(t)} - \frac{{\gamma^{(t)}}}{ {{\bar \gamma }^{(t)}}} \epsilon ( {{{\bm x}^{(t)}};{\bm \omega} } )} \right).
\end{equation}
where $\epsilon \left({\bm x}^{(t)};{\bm \omega} \right)$ predicts the noise added to ${\bm x}^{(t)}$.  From \eqref{eq-29}, the reverse process predicts the Gaussian noise at each step and then removes the predicted noise to restore the ${\bm x}^{(0)}$ from ${\bm x}^{(T)}$ with learnable parameters, which can be viewed the decoding processing.

The loss function for the diffusion-based model at step $t$ is defined as 
\begin{equation}
    {\mathcal L}^{(t)}_{\text{Diff}} = 
    {\mathbb E} \left[ \left\|{\bar {\bm \epsilon}}^{(t)} -  \epsilon \left( {\sqrt{1 - {{\left( {{{\bar \gamma }^{(t)}}} \right)}^2} } {\bm x}^{(0)} +  {{\bar \gamma }^{(t)}} {\bar {\bm \epsilon}}^{(t)};{\bm \omega} , t} \right) \right\|^2 \right].
\end{equation}
During training, we sample the $t$ first and model the ${\bm x}^{(t)}$ with ${\bm x}^{(0)}$ by adding the Gaussian noise with the scheduled variances.

Compared with the previous de-noise frameworks, e.g., DnCNN, that predict the noise with only one step, the de-noising diffusion framework can predict the noise with multiple steps, such that matches the distributions of noise and achieves better performance of de-noise. Therefore, we propose a de-noising diffusion-based signal detection method.

\subsection{The Proposed De-noising Diffusion-based Signal Detection}
The detected signals in \eqref{eq-8} can be rewritten as
\begin{equation}
    {\hat {\bm x}} = {\bm x} + {\tilde {\bm n}},
\end{equation}
where ${\tilde {\bm n}}= \frac{{{{ h}^*}}}{{{{| { h} |}^2}}}{\bm n}$ is an effective noise after the signal detection. We employ the block-fading channel model in \eqref{eq-7}, where the $h$ keeps constant. Therefore, the ${\tilde {\bm n}}$ follows a circularly
symmetric complex Gaussian distribution with zero mean and scaled variance, $\sigma_{\tilde n}^2=\sigma^2_n/|h|^2$.

Since the coefficients of $p\left( {{{\bm x}^{(t)}}\left| {\bm x}^{(t - 1)} \right.} \right)$ should satisfy ${\left( {\sqrt {1 - {\gamma ^{(t)}}} } \right)^2} + {\gamma ^{(t)}} = 1$, we rewritten $\hat {\bm x}$ as
\begin{equation}\label{eq-32}
    {\tilde {\bm x}} = \frac{1}{{\sqrt {1{\rm{ + }}{\sigma_{\tilde n}}} }}{\bm x} + \frac{{{\sigma_{\tilde n}}}}{{\sqrt {1{\rm{ + }}{\sigma_{\tilde n}}} }}{ {\bm \epsilon}},
\end{equation}
where ${\tilde {\bm x}}={\hat {\bm x}}/{\sqrt{1 + \sigma_{\tilde n}}}$ and ${\tilde {\bm n}} = \sigma_{\tilde n}{ {\bm \epsilon}}, { {\bm \epsilon}} \sim {\mathcal{CN}}({\bf 0}, {\bf I})$. 

Comparing \eqref{eq-32} with \eqref{eq-27}, we find that the wireless transmission is similar to the forward process. We model ${\bm x}$ and ${\tilde {\bm x}}$ in \eqref{eq-32} as ${\bm x}^{(0)}$ and ${\bm x}^{(t)}$ in \eqref{eq-27}. It is natural to employ the reverse process to refine $\tilde {\bm x}$, such that obtains the more accurate ${\bm x}$. Given the ${\tilde {\bm x}}$ and ${\sigma_{\tilde n}}$, we adopt \eqref{eq-28} to remove the noise in ${\tilde {\bm x}}$ to closer the ${\bm x}$.  However, the existing variance schedule of $p\left( {{{\bm x}^{(t)}}\left| {\bm x}^{(t - 1)} \right.} \right)$ and sampling algorithm are unsuitable for wireless communications. We need to design the variance schedule and sampling algorithm by considering the channel SNR.

\subsubsection{Variance Schedule Design}
A variance schedule refers to the way in which the mean and variance of the added noise changes over the course of the diffusion process. During this process, the mean and variance of the added noise is adjusted at each step, affecting the amount of noise introduced at each stage, therefore variance schedule determines how the noise level evolves during the diffusion process. 
A variance schedule can impact the quality of generated ${\bm x}$ and the model’s convergence behavior.

The variance schedule should satisfy the ${\bar {\gamma}^{(T)}} \to 0$. Based on the constraint, we design the variance schedule with $T=50$ steps, which is given by
\begin{equation}
    {\gamma^{(t)}} = \frac{0.5t}{T},
\end{equation}
which ${\bar {\gamma}^{(50)}} \approx e^{-6.375} \approx 0 $. The designed variance schedule includes 50 different noise levels. The reasons behind the designed variance schedule can be summarized as
\begin{itemize}
    \item Compared with the conventional diffusion-based framework with 1,000 steps for generative tasks, we empirically find that the de-noise task does not need too many steps due to the low complexity of the de-noise task.
    \item We design a monotonic function of ${\gamma^{(t)}}$ to achieve coarse-to-fine de-noise processing, which has an unequal interval SNR, e.g., a small interval in high SNR regions and a large interval in low SNR regions. The unequal interval SNR can speed up the de-noise processing with fewer steps at low SNR regions.
\end{itemize}

\subsubsection{Sampling Algorithm}
The sampling algorithm performs the reverse process by sampling the steps. For example, the conventional diffusion-based framework usually samples 1,000 steps from $T\to 0$ \cite{HoJA20} or 100 steps with the subsequence of $T\to 0$ \cite{ SongME21}, in which ${\bm x}^{(T)}$ is the first sampled step. However, starting from ${\bm x}^{(T)}$ is unsuitable for signal detection.  The detected signals will start from different ${\bm x}^{(t)}$ where $t$ depends on the received SNR at the receiver. Therefore, we propose a dynamic sampling algorithm shown in Algorithm \ref{alg-4}. 

Firstly, given the known ${\sigma_{\tilde n}}$, we search the starting point $\tilde t$ at the reverse process, which is given by
\begin{equation}\label{eq-34}
    \frac{1}{\sqrt{1 + \sigma_{\tilde n}}} \in [{\bar \gamma}^{({\tilde t}+1)}, {\bar \gamma}^{({\tilde t})}].
\end{equation}

Then, the signal detection aims to recover the transmitted signals as more accurate as possible. Therefore, we change the random sampling to deterministic sampling. In detail, we reduce the degree of randomness in the reverse process by setting $\sigma^{(t)}$ in \eqref{eq-28} equals to zero, which means that the $q\left( { {\bm x}^{(t - 1)}\left| {{{\bm x}^{(t)}}} \right.} \right)$ changes from $ {\mathcal N}\left( { {\mu \left( { {\bm x}^{(t)}; {\bm \omega} } \right), \sigma^{(t)} {\bf I}}} \right)$ to deterministic $\mu \left( { {\bm x}^{(t)}; {\bm \omega} } \right)$.

\begin{algorithm}[!t]
\caption{Dynamic Sampling Algorithm.}
\label{alg-4}
\SetKwInput{KwInput}{Input}                
\SetKwInput{KwInitia}{Initialization}
\SetKwInput{KwOutput}{Output}              
\SetKwInput{KwRet}{Return}
\DontPrintSemicolon

\SetKwFunction{FMain}{Main}
\SetKwFunction{FCC}{Dynamic Sampling}

  \SetKwProg{Fn}{Function}{:}{}
  \Fn{\FCC{}}{
        \KwInput{The detected signal ${\tilde {\bm x}}$ and $\sigma_{\tilde n}$.}
             Initialize ${\tilde {\bm x}}$ as the starting point, ${\bm x}^{(\tilde t)}$. \;
             Find the $\tilde t$ by \eqref{eq-34} with $\sigma_{\tilde n}$. \;
            \For{$t = \tilde t \to 1$ }
             {${\bm x}^{(t-1)} = \frac{1}{{\sqrt {1 - {\gamma^{(t)}}} }}\left( {{\bm x}^{(t)} - \frac{{\gamma^{(t)}}}{ {{\bar \gamma }^{(t)}}} \epsilon ( {{{\bm x}^{(t)}};{\bm \omega} } )} \right)$ \;
            }
        \KwRet{${\bm x}^{(0)}$}
  }

\end{algorithm}

\section{Numerical Results}\label{sec-v}

In this section, we compare the proposed HDA-DeepSC with DL-based semantic communication systems and digital communication systems over AWGN and Rician fading channels, where we assume the perfect CSI for all schemes.  

\subsection{Implementation Details}
\subsubsection{The Dataset}
We choose the \textit{DIK2K} dataset \cite{Ignatov_2018_ECCV_Workshops} for training, which contains 1,000 images with different scenes. The \textit{Kodak} dataset is used for testing.

\subsubsection{Training Settings}
The semantic codec consists of 6 Swin-Transformer layers, respectively. Each layer is with 6 heads and a width of 120. The diffusion-based model adopts the structures of OpenAI-UNet. The ${\lambda}_{\mathcal F}$, $\lambda_z$, and $\lambda_r$ is 0.1, 0.1, and 0.0005, respectively. The learning rate is $2\times 10^{-4}$. 

\subsubsection{Benchmarks and Performance Metrics}
We adopt the separate source-channel coding, the DL-based analog semantic communication system, the DL-based digital semantic communication system, and the one-step denoising network as the benchmarks, which are detailed as follows.
\begin{itemize}
    \item Separate source-channel coding: Employ the source and channel coding separately to transmit the images, we use the following technologies, respectively:
    \begin{itemize}
        \item Better Portable Graphics (BPG) for image source coding, the state-of-the-art image compression method.
        \item Low-density parity check (LDPC) coding and capacity-achieved coding are used for the channel coding. 
        \item The adaptive modulation and coding (AMC) is employed for different SNRs, including 1/2 coding rate with BPSK, 1/2 coding rate with QPSK, 3/4 coding rate with QPSK, 1/2 coding rate with 16QAM, and 3/4 coding rate with 16QAM.
    \end{itemize}
    \item Analog semantic communication systems: The purely analog semantic communication of HDA-DeepSC trained with MSE loss.
    \item Digital semantic communication systems: The DeepJSCC-Q proposed in \cite{Tung9998051}.
    \item Denoising convolutional neural network (DnCNN) as the one-step de-noise benchmark 
\end{itemize}

\begin{figure}[!t]
    \centering
    \includegraphics[width=80mm]{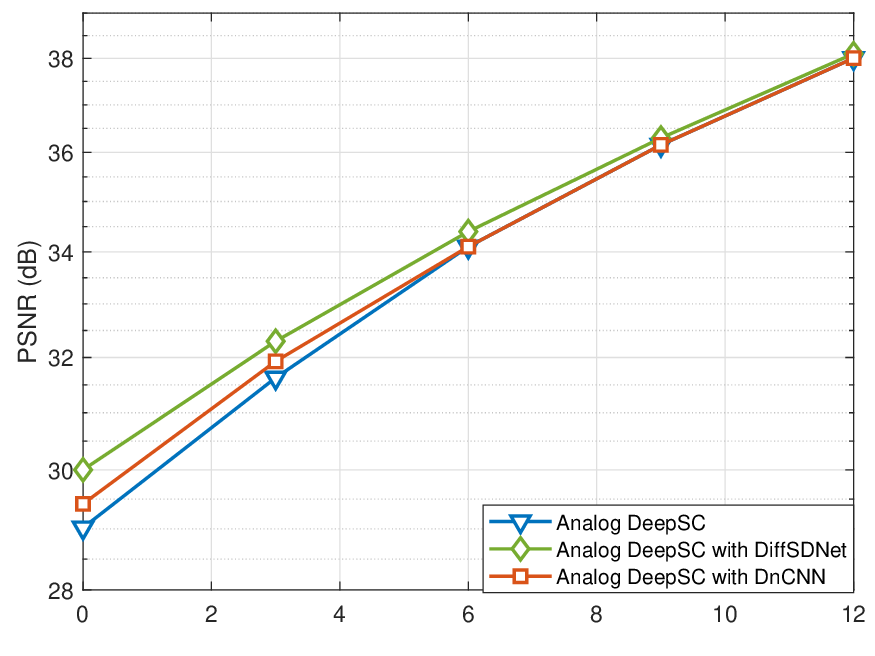}
    \caption{{The PSNR performance comparison between Analog DeepSC and Analog DeepSC with different denoisers on the Kodak dataset.}}
    \label{fig:denoising}
\end{figure} 

\begin{table}[!t]
\caption{The PSNR Comparison between the analog DeepSC with different diffusion-based denoisers at SNR=0dB.}
\label{tab:1}
\centering
\begin{tabular}{ c|c|c|c|c|c } 
\toprule
Sampling Steps& 10  & 20 & 30 & 40 & 50  \\
\midrule
DDPM & 29.3 & 29.0 & 28.9 & 28.8 & 29.0  \\
\midrule
DiffSDNet & 29.1 & 29.5 & 29.9 & 30.1 & $\backslash$     \\
\bottomrule
\end{tabular}
\end{table}

The LDPC codes we use are from the 802.11ad standard, with blocklength 672 bits for both the 1/2 and 3/4 rate codes. The coherent time is set as the transmission time for each image in the simulation. We set $r = 1$ for the Rician channels and $h = 1$ for the AWGN channels. Peak signal-to-noise ratio (PSNR) and multi-scale structural similarity (MS-SSIM) are used as the metrics to measure the local and global quality of images.

\subsection{Denoising Networks Comparisons}
Fig. \ref{fig:denoising} presents the PSNR performance for the analog DeepSC with different denoisers. First observe that the analog DeepSC with denoiser has a larger PSNR than that without denoiser in the low SNR regimes. This validates the effectiveness of the denoiser in reducing the noise level. For the small noise level at the high SNR regimes, the analog DeepSC is capable of restoring the signals therefore all methods achieve a similar PSNR as the SNR increases. Furthermore, we observe that the analog DeepSC with DiffSDNet outperforms that with DnCNN with 0.6dB in terms of PSNR. This suggests that the multiple-step denoiser has a stronger power of denoising than the one-step denoiser. 

Table \ref{tab:1} shows the comparison between analog DeepSC with DDPM and DiffSDNet. The proposed DiffSDNet can achieve higher PSNR with fewer sampling steps than the DDPM, confirming the effectiveness of the designed variance schedule and sampling algorithm. Especially, the PSNR of analog DeepSC with DDPM will decrease as the number of sampling steps increases. This is due to the high degree of randomness introduced in the reverse process.

\begin{figure*}[!t]
	\centering
	\subfigure[PSNR]{
			\includegraphics[width=80mm]{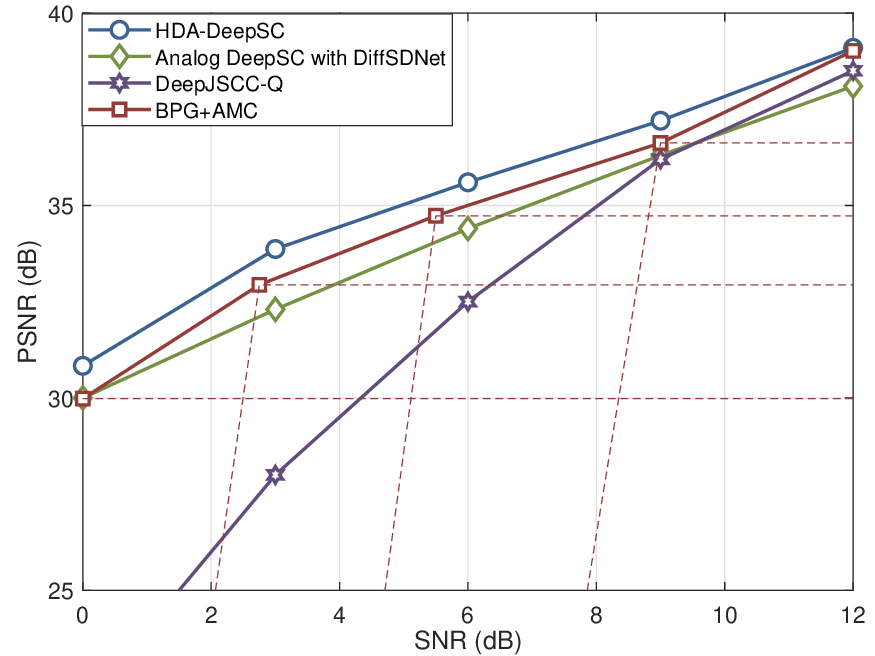}
		\label{fig:agwn-psnr}
	}
    	\subfigure[MS-SSIM]{
		 	\includegraphics[width=80mm]{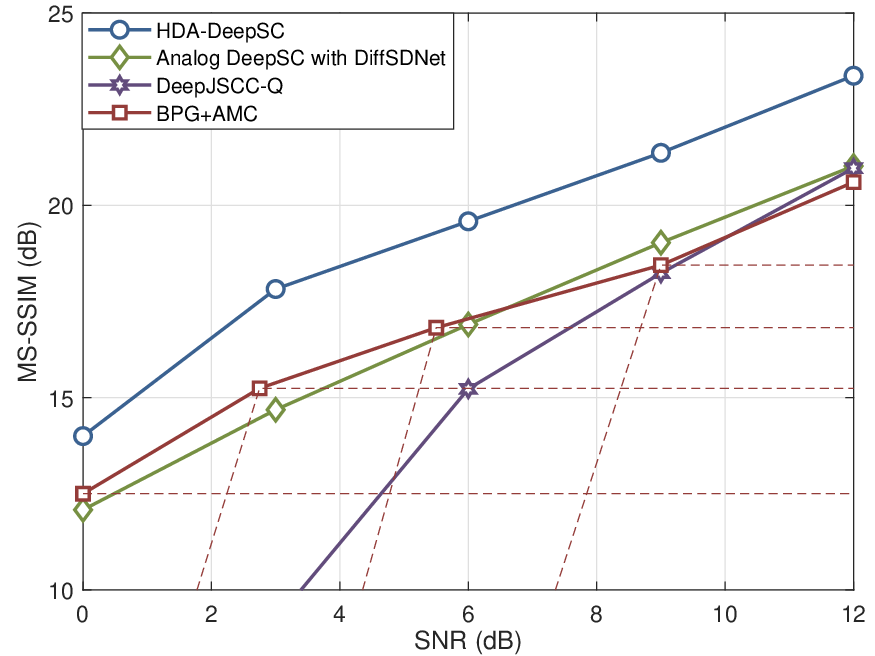}
		\label{fig:awgn-msssim}
    }
    
	\caption{Comparison between HDA-DeepSC and the Analog DeepSC, DeepJSCC-Q, and BPG with different channel coding on the Kodak dataset over AWGN channels.}
	\label{fig:awgn}
\end{figure*}
\begin{figure*}[!t]
	\centering
	\subfigure[PSNR]{
			\includegraphics[width=80mm]{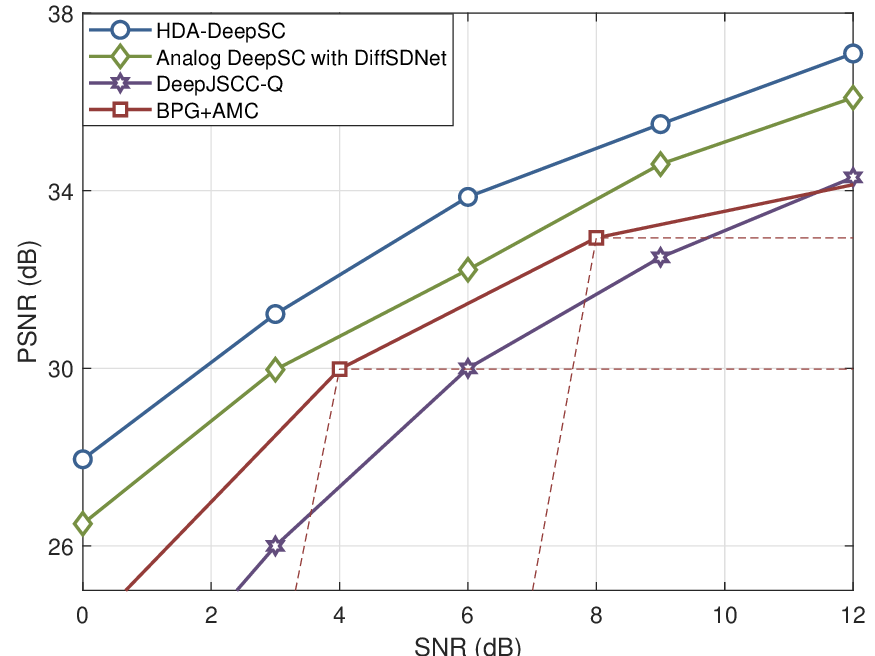}
		\label{fig:rician-psnr}
	}
    	\subfigure[MS-SSIM]{
		 	\includegraphics[width=80mm]{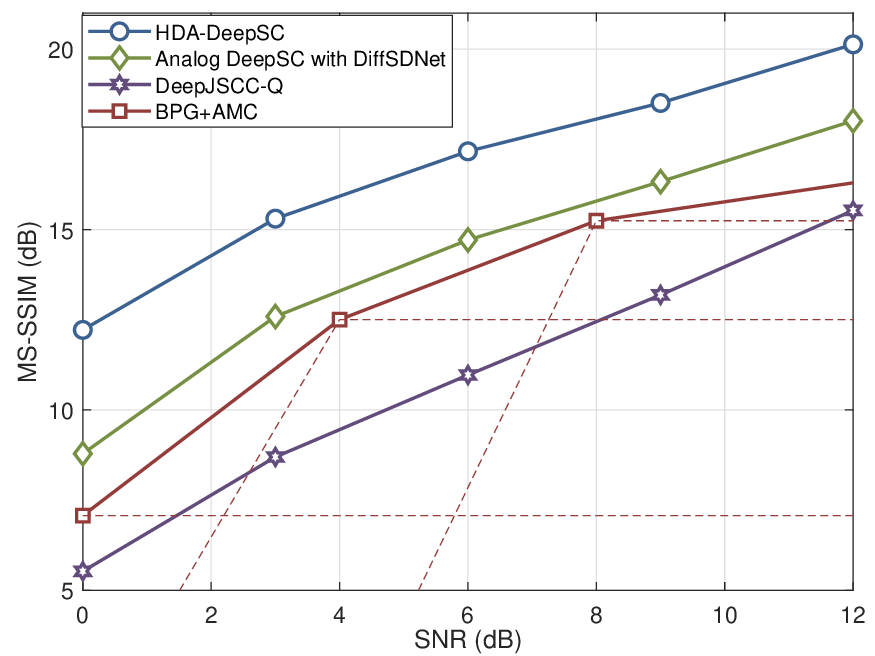}
		\label{fig:rician-msssim}
    }
    
	\caption{Comparison between HDA-DeepSC and the Analog DeepSC, DeepJSCC-Q, and BPG with different channel coding on the Kodak dataset over Rician channels.}
	\label{fig:rician}
\end{figure*}

\subsection{Communication System Comparisons}

\begin{figure*}[!t]
    \centering
    \includegraphics[width=180mm]{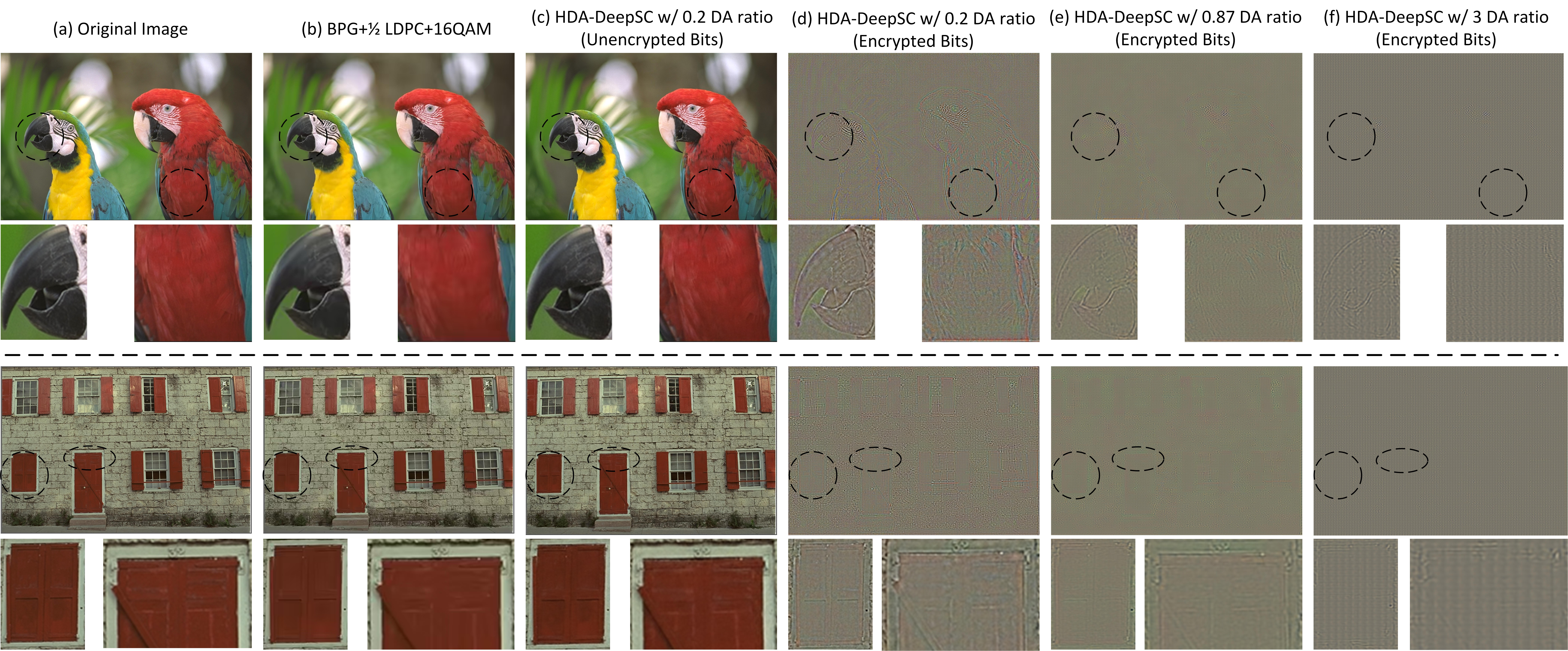}
    \caption{Visualized examples for different methods transmitted over AWGN channels at SNR=10dB: (a) original image; (b) image recovered by BPG with 1/2 LDPC and 16QAM; (c) image recovered by HDA-DeepSC with 0.2 DA ratio using unencrypted bits; (d)-(f) image recovered by HDA-DeepSC with 0.2, 0.87, and 3 DA ratio using encrypted bits, respectively.}
    \label{fig:visualized-results}
\end{figure*}

Figs. \ref{fig:awgn} and \ref{fig:rician} report the PSNR and MS-SSIM comparison between the various methods over AWGN channels and Rician channels. For AWGN channels, we can see in Fig. \ref{fig:awgn} that our HDA-DeepSC outperforms all the benchmarks. This indicates that the discrete signals of the digital component can accurately deliver crucial semantic information for details recovery and the continuous signals of the analog component can prevent the leveling-off and cliff-edge effects for lower quantization errors. Besides, the HDA-DeepSC achieves the best performance in terms of MS-SSIM, which means that the images transmitted by HDA-DeepSC have better global quality. This is likely because we introduce the Fourier-based loss function that makes the model learn the long-distance dependencies.  For the Rician channel case shown in Fig. \ref{fig:rician}, we observe that the DL-based analog systems are more robust to channel changes due to the high degree of freedom in continuous signals, in which the HDA-DeepSC is beneficial from the analog component. Visual examples are presented in Fig. \ref{fig:visualized-results}(a)-(c). We can observe the proposed HDA-DeepSC can restore more details, e.g., the mouth and feathers of the parrot, than the BPG with LDPC and 16QAM due to delivering essential semantic information accurately by the digital transmitter.

\begin{table}[!t]
\caption{The Ablations of Fourier-based Component: MSE Loss, MSE Loss with Fourier-based Module, and MSE Loss with Fourier-based Loss.}
\label{tab:2}
\centering
\begin{tabular}{ c|c|c|c } 
\toprule
Method & MSE loss  & \makecell[c]{MSE loss + \\ Fourier-based Module}  &  \makecell[c]{MSE loss + \\ Fourier-based loss} \\
\midrule
PSNR & 48.09  & 50.35 & 52.19    \\
\midrule
MS-SSIM & 33.01 & 35.23 & 40.01    \\
\bottomrule
\end{tabular}
\end{table}

In Table \ref{tab:2}, we study the ablations of Fourier-based components by only considering the semantic codec, in which the MSE loss with Fourier-based module means that we insert the pluggable Fourier-based modules \cite{ChiJM20} into the semantic codec and train the semantic codec with MSE loss function. The Fourier-based module or loss can improve the quality of images with more than 2dB in terms of PSNR and MS-SSIM due to the long-distance dependencies learning in the frequency domain. Besides, we observe that the Fourier-based loss can largely increase MS-SSIM than the Fourier-based module. The reason behind that is the Fourier-based module introduces the additional Fourier-based parameters making it challenging to further improve its performance. This suggests that Fourier-based loss can directly capture the global information of images without additional parameters and hence as an attractive loss to improve the global quality of images. 

\begin{figure*}[!t]
	\centering
	\subfigure[PSNR]{
			\includegraphics[width=80mm]{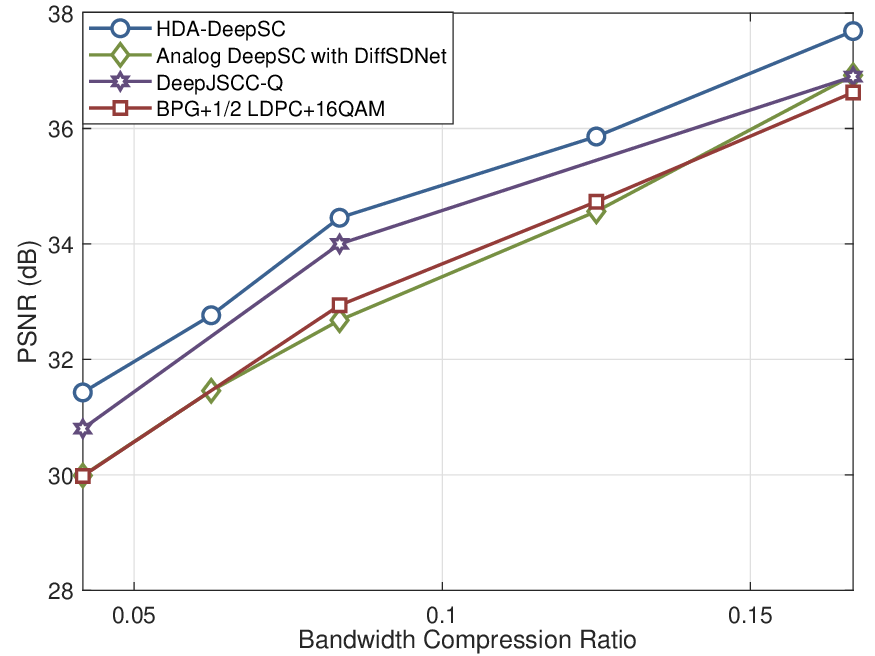}
		\label{fig:bandwidth-psnr}
	}
    	\subfigure[MS-SSIM]{
		 	\includegraphics[width=80mm]{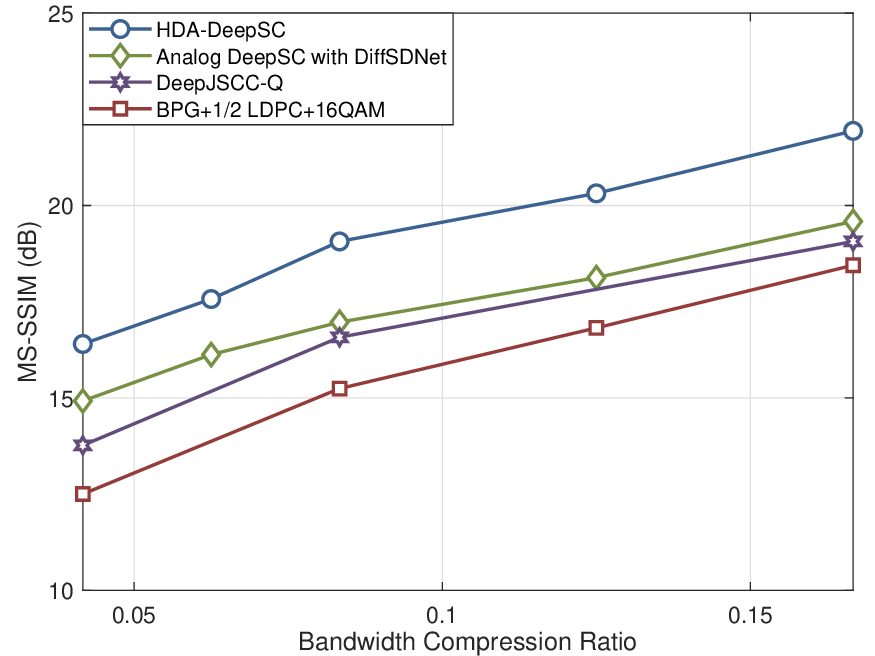}
		\label{fig:bandwidth-msssim}
    }
    
	\caption{PSNR and MS-SSIM performance for different bandwidth compression ratios on the Kodak dataset over AWGN channels.}
	\label{fig:bandwidth}
\end{figure*}

\begin{figure*}[!t]
	\centering
	\subfigure[PSNR]{
			\includegraphics[width=80mm]{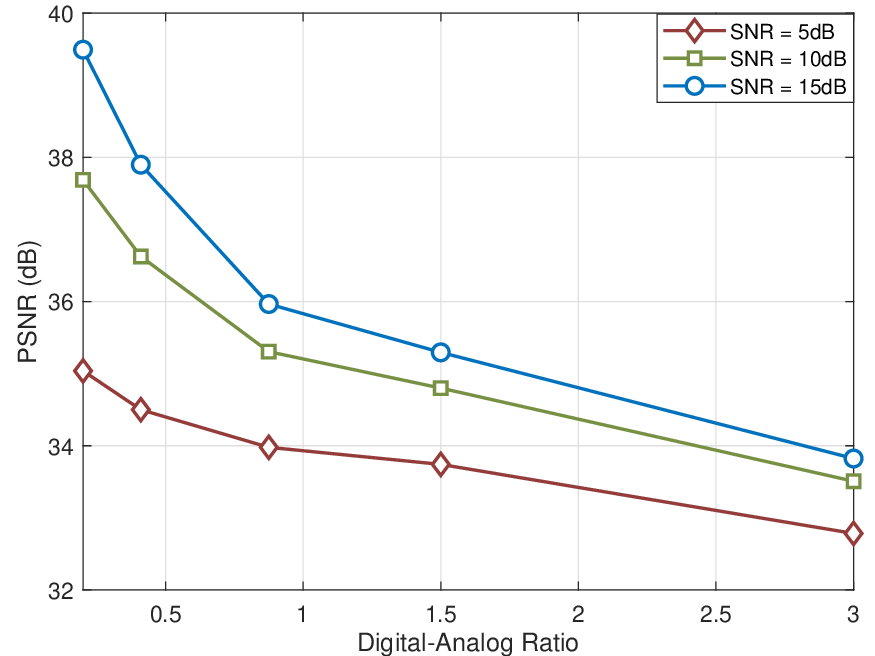}
		\label{fig:adratio-psnr}
	}
    	\subfigure[MS-SSIM]{
		 	\includegraphics[width=80mm]{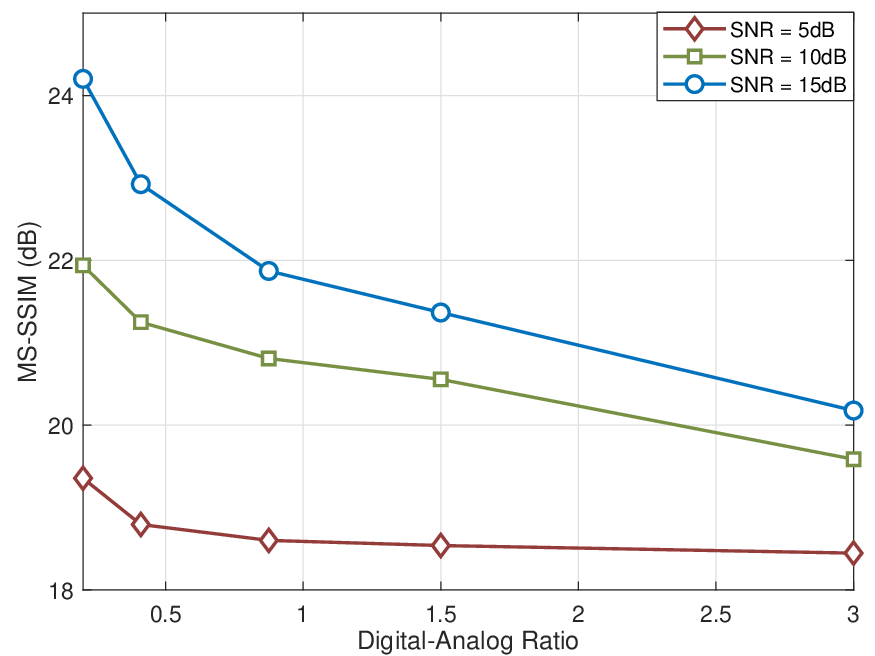}
		\label{fig:adratio-msssim}
    }
    
	\caption{PSNR and MS-SSIM performance for different digital-analog ratios on the Kodak dataset over AWGN channels.}
	\label{fig:adratio}
\end{figure*}

\subsection{Bandwidth Compression Ratio Comparisons}
Fig. \ref{fig:bandwidth} demonstrates the comparisons for different bandwidth compression ratios over AWGN channels at SNR=10dB. The HDA-DeepSC outperforms all the benchmarks in terms of PSNR and MS-SSIM. For example, the HDA-DeepSC achieves the same PSNR as separate codings (the BPG with 1/2 LDPC and 16QAM) with a 33\% improvement on bandwidth compression ratio. This suggests that the HDA-DeepSC can provide a higher data transmission rate than the benchmarks for a given PSNR or MS-SSIM. Besides, we find that the learning-based methods outperform the BPG in terms of MS-SSIM, indicating the neural networks operate as the better content generator, thereby generating the image with global consistency.

\subsection{Digital-Analog Ratio Comparisons}

Fig. \ref{fig:adratio} shows the comparisons across different digital-analog (DA) ratios by changing the ratio between the number of transmitted symbols of digital and analog components, where the total number of transmitted symbols is fixed. The larger DA ratio means more semantic information is transmitted with the digital transmitter and vice versa. We can observe that the PSNR and MS-SSIM decrease as the DA ratio increases, which is caused by the unavoidable quantization errors introduced by the digital transmitter. The more semantic information transmitted through the digital transmitter, the larger the quantization errors introduced to the transmitted information. This suggests that the analog transmitter operates as a continuous signal-based system, thereby effectively reducing the quantization errors by decreasing the DA ratio. 

\begin{table}[!t]
\caption{The PSNR Performance for the Encrypted and Unencrypted Bits over AWGN Channels at SNR=10dB. }
\label{tab:3}
\centering
\begin{tabular}{ c|c|c|c|c|c } 
\toprule
DA Ratio & 0.2  & 0.4 & 0.875 & 1.5 & 3  \\
\midrule
Encrypted Bits & 13.75 & 13.67 & 13.61 & 13.56 & 13.52  \\
\midrule
Unencrypted Bits & 37.69 & 36.62 & 35.30 & 34.80 & 33.51    \\
\bottomrule
\end{tabular}
\end{table}

\subsection{Data Security}
Table \ref{tab:3} reports the PSNR performance for the encrypted and unencrypted bits, where these terms refer to whether the encryption algorithm encrypts the bit streams transmitted by the digital transmitter. We assume that the eavesdropper is incapable of decoding the encrypted bits and only decodes the semantic information transmitted by the analog transmitter, where the HDA-DeepSC model is known to the eavesdropper. From Table \ref{tab:3}, the PSNR of encrypted bits is 20dB lower compared to that of unencrypted bits, indicating the images recovered by encrypted bits are little like the original ones. In other words, the eavesdropper obtains less information from the semantic information transmitted by the analog transmitter. Besides, the PSNR of encrypted bits slightly decreases as the DA ratio increases. This suggests that the HDA-DeepSC effectively safeguards data with few bits while achieving the high PSNR. Visual examples are presented in Figs. \ref{fig:visualized-results}(c)-(f), where Figs. \ref{fig:visualized-results}(d)-(f) are the images recovered by encrypted bits. Interestingly, the essential information is protected by the HDA-DeepSC, e.g. the color, the background, and the textures, which proves the effectiveness of the HDA-DeepSC in data security.

\section{Conclusion}\label{sec-vi}

In this paper, we present an innovative HDA semantic communication framework that combines the strengths of analog and digital semantic communications. Our framework aims to overcome the inherent limitations associated with each approach. Building upon the framework, we introduced a robust HDA semantic communication system called HDA-DeepSC, specifically designed for multimedia transmission. HDA-DeepSC leverages digital communication methods to transmit crucial semantic information, ensuring accurate delivery and data security. Additionally, it utilizes analog communication methods to transmit auxiliary semantic information, effectively mitigating the leveling-off and cliff-edge effects associated with traditional approaches. We also introduced analog-digital allocation and fusion modules to separate and fuse the digital and analog components, respectively. Besides, we have designed the Fourier-based loss function to guide the model in learning the long-distance dependencies and combined the rate constraint with the non-parametric, fully factorized density model. Moreover, we have proposed the diffusion framework enhanced signal detection, named DiffSDNet, by multiple denoising steps to reduce the noise level at the low SNR regimes, in which we customized the variance schedule and sampling algorithm for wireless communication environments. The numerical results have proved the effectiveness of DiffSDNet in denoising and demonstrated the superiority of HDA-DeepSC in terms of robustness, transmission rate, and data security, especially in low SNR regimes. Therefore, the proposed HDA semantic communication framework shows great promise as a candidate for the new semantic communication paradigm, offering significant potential for real-world implementations.

\ifCLASSOPTIONcaptionsoff
  \newpage
\fi

\bibliographystyle{IEEEtran}
\bibliography{reference.bib}

\end{document}